\documentclass[twocolumn,showpacs,preprintnumbers,amsmath,amssymb,revsymb,prb, superscriptaddress]{revtex4}
\usepackage[dvips]{graphicx}
\usepackage{mathptmx}

\newcommand{\D}[2]{\frac{d #1}{d #2}}

\newcommand{\DD}[2]{\frac{d^2 #1}{{d #2}^2}}

\begin{document}
\title{Edge-induced spin polarization in two-dimensional electron gas}
\author{P. Bokes } \email{peter.bokes@stuba.sk}
\affiliation{Department of Physics, Faculty of Electrical Engineering and
        Information Technology, Slovak University of Technology,
    Ilkovi\v{c}ova 3, 812 19 Bratislava, Slovak Republic}
\affiliation{European Theoretical Spectroscopical Facility (ETSF, {\tt www.etsf.eu})}
\author{F. Horv\'{a}th}
\affiliation{Department of Physics, Faculty of Electrical Engineering and
        Information Technology, Slovak University of Technology,
    Ilkovi\v{c}ova 3, 812 19 Bratislava, Slovak Republic}

\date{\today{}}

\begin{abstract}
We characterize the role of the spin-orbit coupling between electrons and the confining potential 
of the edge in nonequilibrium 2D homogeneous electronic gas. 
We derive a simple analytical result for the magnitude of the current induced spin polarization 
at the edge and prove that it is independent of the details of the confinement edge potential and the 
electronic density within realistic values of the parameters of the considered models. 
While the amplitude of the spin accumulation is comparable to the experimental values of extrinsic 
spin Hall effect in similar samples, the spatial extent of edge induced effect is restricted 
to the distances of the order of Fermi wavelength ($\sim$ 10nm). 

\end{abstract}


\maketitle

\section{Introduction}
\label{sec-1}
One of the exciting new discoveries in solid state physics in the 
last few years has been the experimental observation of the extrinsic spin Hall
effect\cite{Kato04,Sih05,Wunderlich05,Sih05b} in GaAs heterostructures. 
The mechanism of this effect relies on the  spin-orbit coupling between the spin of electrons 
with the perturbing potential of impurities. Another example of similar coupling is the
Rashba-Bytchkov interaction in asymmetrically doped GaAs quantum wells containing two-dimensional 
electron gas (2DEG) where the coupled potential comes from the internal electrostatic electric 
field induced by the structural asymmetry of the well. In modeling both of these situations, 
the atomic potential of the ideal bulk semiconductor enters only through the renormalization 
of the effective mass and the spin-orbit interaction strength.  

Interestingly, the Rashba-Bytchkov interaction has been also suggested as a source of spin Hall-like 
phenomenology\cite{Sinova04} for a clean 2DEG. While the effect 
has been shown to disappear in extended 2D systems with arbitrarily weak 
disorder\cite{Inoue04,Rashba04}, it is now known that finite size of the sample in combination with 
Rashba-Bytchkov coupling results in the transverse spin current and accumulation of opposite spin 
densities at the edges of the sample\cite{Nikolic05a,Moca07}. Furthermore, it has been found that 
boundaries might directly influence the spin polarization due to the Rashba-Bytchov coupling 
either in very wide 2D electronic systems with an edge\cite{Usaj05,Reynoso06,Zyuzin07,Teodorescu09,Sonin09c} 
or narrow Rashba strips\cite{Yao06,Li08}. However, the physical origin of the latter is quite 
different from the former: whereas the bulk Rashba-Bytchkov effect arises from accelerations 
of electrons in the external electric field, the latter result from the reflection of electrons from the 
sample's boundary.

Motivated by this development, it is natural to explore further alterations 
of the perfectly periodic bulk potential and its consequences 
on the spin polarization in the presence of the current. Several authors 
have considered such a situation in quantum wires with a parabolic confining 
potential\cite{Bellucci06,Hattori06, Xing06,Bellucci07}, wider strips with 
parabolic\cite{Jiang06} or abrupt\cite{Bokes08} confinement edge. These alterations 
in the potential landscape can be also viewed as the ``impurity'' which leads to nonzero 
spin polarization induced by the flow of the electric current. Recently, experimental evidence 
of this effect of the in-plane field on spin polarization has been reported\cite{Debray09}. 
Further enhancement of this kind of spin-polarization has been suggested using a transport through chaotic 
quantum dot\cite{Krich08} or resonant tunable scattering center\cite{Yokoyama09}.

In our previous work\cite{Bokes08} we have compared the magnitude of the spin polarization 
induced by the latter mechanism with the one induced by the Rashba-Bytchkov coupling and scattering of the 
edge\cite{Zyuzin07}. It turns out that for a typical 2DEG with an edge,
it is about three orders of magnitude larger than the one caused by the edge and the Rashba-Bytchkov 
interaction. This substantial difference is mostly due to the fact that the parameter characterizing 
spin-orbit coupling appears in second order in the case of the Rashba-Bytchov-based mechanism\cite{Zyuzin07} 
whereas for the edge-induced spin-orbit coupling it comes already in the first order.

In this paper we consider the edge of the 2DEG in Si doped GaAs 
quantum well, a system for which the spin Hall effect has been
experimentally observed. This allows us to estimate the values of all 
the parameters entering and we can  compare the importance of this edge-effect 
to the contribution of other mechanisms leading to current-induced spin polarization.
The models considered here are reasonably realistic yet simple enough 
to be solved almost analytically. We derive a generally valid simple analytical formula 
for the induced number of electrons with un-compensated spin per unit length of the edge,
\begin{equation} 
	m_z = - \frac{m_e}{\hbar q_e} \alpha_E j, \label{eq-m_z-SI}
\end{equation}
where $j$ is the electrical current density in the 2DEG, $\alpha_E$ is the parameter 
giving the strength of the spin-orbit coupling~\cite{Engel05},
and $m_e$ and $q_e$ are the effective mass and the charge of electron respectively. Noticeably, 
this result is independent of the density of the 2DEG and the details of the edge potential 
as long as the electrons are confined within the sample.

\section{The model of the edge and spin-orbit coupling}
\label{sec-2} 

\begin{figure}
\begin{center}
      \includegraphics{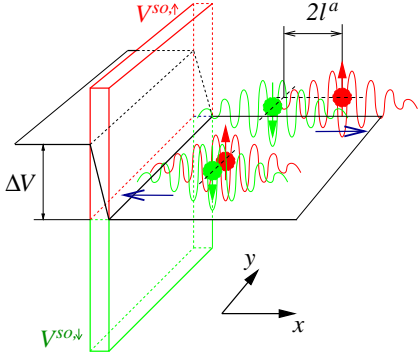}
\caption{(Color online) The edge's confinement potential of depth $\Delta V$ leads, via the SO coupling 
	(Eq.~\ref{eq-V-SO-2D}) to attractive potential ($V^{SO,\downarrow}$, in green) for electrons 
	with spin down in the vicinity of the edge. The initially compensated spins of electrons approaching 
	the edge becomes locally un-compensated due to the difference in the scattering shift $l^a$.
	} \label{fig-1}
\end{center}
\end{figure}

Let us consider a sample with 2DEG created within a GaAs-based quantum well. The electrical current in the 
plane of the 2DEG will be driven along the $y$ direction. We will be interested in the physics close 
to one of the two edges of the sample along which the current flows. The coordinate in the direction 
perpendicular to the edge will be $x$, $x>0$ corresponding to the region where the 2DEG is present 
(see Fig.~\ref{fig-1}).

We will describe the electronic states for electrons in the 2DEG within the effective mass 
approximation, treating the electrons as noninteracting quasiparticles. Qualitative changes 
in our results introduced within a self-consistent mean-field treatment will be studied afterwards.
The length scale characterizing the electrons is given by their Fermi wavelength.  
It typically attains values\footnote{We will use effective atomic units; the distances are measured in 
multiplies of the effective Bohr radius, $a_B^*=\frac{\epsilon_r}{m_{ef}}=9.79$nm, energy 
in the effective Hartrees, {\it Ha}$^*=\frac{m_{ef}}{\epsilon_r^2} \textrm{\it Ha}=11.9$meV, 
both numerical values are given for GaAs where $m_e^*=0.067m_e$ and $\epsilon_r=12.4$ are the effective 
mass of the electron and the relative permittivity.} 
$\lambda_F = \sqrt{2\pi/n_{2D}} \sim 2.5 a_B^*$, where $n_{2D}$ is the two-dimensional electronic
density and $a_B^*=9.79$nm is the effective Bohr radius in GaAs. Since this is about two orders 
of magnitude larger than the inter-atomic distances, we can employ the effective mass theory and 
approximate the form of the confining potential with a simple functional form. 

In our work will assume that the confining edge potential, $V(x)$, is independent of the coordinate $y$,
directed along the edge.
Within our analytical derivations we model $V(x)$ as an abrupt step of height $\Delta V$,
$V(x) = V_\theta(x) = \Delta V \theta(-x)$, where $\theta(x)$ is the unit step function. 
The actual value of the step is much larger than the Fermi energy of the 2DEG. This is frequently 
used to set the potential step to infinity. However, here it is essential that the step 
is finite as the whole spin-orbit coupling is nonzero only in the region of the nonzero gradient 
of the confining potential.
Since this value should be of the order of the work function of the electrons in the 2DEG,
we fix this value to $\Delta V = 2.7$eV$\sim 230$Ha$^*$. One of the results of our work is 
the demonstration that the current induced spin polarization is rather independent of this value so that 
we do not need to be concerned with its precise numerical value, as long as it is large ($>>E_F$)
but finite value. 

Once we establish the analytical result for this model of abrupt potential step, we will also consider 
more general forms: step with a linear slope on a distance $d$, $V_d(x)$ (used in Fig.~\ref{fig-1}),  
and a partially self-consistent, density-dependent model with a small triangular-shaped dip, 
$V_n(x)$, close to the edge. In the absence of this dip, the decrease in the density to zero at the edge in the 2DEG results in 
a slight local depletion of electrons there. A small negative value of the depth of the dip, $-V_0$, 
that is found self-consistently results in a charge neutral edge which is the physically expected 
situation.

The essential part of our model is the spin-orbit (SO) coupling term in the electrons' Hamiltonian. 
In general, several different types of SO interactions can be similar in magnitude: the 
Rashba-Bytchkov, the Dresselhaus or the impurity-induced SO coupling. However, since they are all 
small perturbations to the effective Hamiltonian, we can consider them separably and superpose 
their outcomes in the sense of first order perturbation theory in terms of SO parameter
appearing in the SO interaction. In our paper we will consider a special case of the impurity-induced 
SO coupling for the special case when the ``impurity'' is the edge potential only.
In general, the SO contribution to the Hamiltonian takes the form\cite{WinklerBook}
\begin{eqnarray}
	\hat{V}^{SO} = \alpha_E \vec{\sigma} \cdot \left( \vec k \times \nabla V(x)  \right),
	\label{eq-SO-vector}
\end{eqnarray}
where $\alpha_E$ characterizes the strength of the SO coupling, $\vec{\sigma}$ are Pauli matrices, 
$\vec k$ is the operator of electron's momentum and $V(x)$ is the effective one-electron potential 
energy. Within our study the latter corresponds to various models of the edge potential: 
$V_\theta(x)$, $V_d(x)$ or $V_n(x)$. The strength of the SO coupling in GaAs is known to be 
approximately $\alpha_E \approx 5.3 \AA^2 = 5.53 \times 10^{-4} $a$_B^*$ 
which is $10^6$ times larger than the strength of the SO coupling in vacuum\cite{Engel05}.

Within the framework of this model it is very easy to understand the origin of the current-induced 
spin polarization. The gradient of the potential energy is directed only in the $x$ direction,
the momentum is confined only within the $xy$ plane so that the only nonzero component 
of the spin operator comes from its $z$ component due to the mixed product form in Eq.~\ref{eq-SO-vector}. 
Specifically, the SO interaction is simply a spin-dependent potential 
\begin{equation}
	\hat{V}^{SO} = - \alpha_E \sigma_z q V'(x) \label{eq-V-SO-2D}
\end{equation}
with a well defined quantum number for the projection of the spin in the $z$ direction, $\sigma_z$, 
attaining values $\pm 1$, and $q$ is the $y$ component of the electron's 
momentum (Fig.~\ref{fig-1}). Hence, the spin-orbit term is a potential energy which, 
for states with positive momentum
in the $+y$ direction (the current), is smaller (larger) in the spatial region of 
negative derivative of the edge potential, $V'(x)<0$, for electrons with their spin down (up).
Since close to $x=0$ we have $V'(x) =  - \Delta V \delta(x) < 0$, we expect that close to the edge 
we will find majority of spin-down electrons. Of course, in equilibrium for zero total current density 
there will be equal number of electron with $q>0$ and $q<0$ so that in this special case 
zero total spin polarization will be observed.

\section{Spin polarization and the spin-dependent phase-shift}
\label{sec-3}

In our previous work~\cite{Bokes08}, we have given estimates for the total induced spin per unit 
length of the edge using arguments based on wavepacket propagation. We considered a 
single-electron wavepacket state, $\psi^{q,\sigma}_{k,\Delta k}(t)$, for a specific momentum 
in the $y$ direction $q$, spin $\sigma$, with momentum $k$ of uncertainty $\Delta k$, 
\begin{eqnarray} 
	\psi^{q,\sigma}_{k,\Delta k}(t) &=& \int_{k-\Delta k/2}^{k+\Delta k/2} \frac{dk}{\sqrt{2\pi \Delta k}}
			\left(e^{-ikx} + r^{q,\sigma} (k) e^{ikx} e^{iqy}
			\right) \nonumber \\ && \times e^{-i(k^2/2 + q^2/2) t },
\end{eqnarray}
where $r^{q,\sigma}(k)$ is the reflection amplitude of the eigenstate with energy $E=(1/2)(k^2+q^2)$.
For large negative times only the incoming part (moving from the right in the Fig.~\ref{fig-1}) 
of this wavepacket will give rise to nonzero contributions (the reflected part will be extremely small 
due to the quickly oscillating factors) and the wavepacket will be approaching 
the edge (Fig.~\ref{fig-1}, moving to the left). 
On the other hand, for large positive times only the reflected wave will give nonzero contribution. 
The position and form of the reflected wavepacket will depend on the spin of the incoming wavepacket since 
the reflection amplitude, $r^{q,\sigma}(k)$ will be different for the two possible spin orientations. 
Specifically, it is shown in the Appendix A that the reflection amplitude has the form 
$r^{q,\sigma}(k) = e^{i\theta^{q,\sigma}_k}$ where the phase shift up to 3rd order in $\alpha_E$ is
\begin{eqnarray}
 	\theta^{q,\uparrow/\downarrow}_k &=& \pi + 2 \textrm{atan}\frac{k}{\kappa} 
			\mp 2 k \alpha_E q + 2 k \kappa \alpha_E^2 q^2  \nonumber \\
		&&	\mp 4 \alpha_E^3 q^3 k (\Delta V - \frac{2}{3} k^2),  \label{eq-phase-shift}
\end{eqnarray}
where $\kappa=\sqrt{2 \Delta V - k^2}$. Assuming the width $\Delta k$ of the wavepacket 
$\psi^{q,\sigma}_{k,\Delta k}(t)$ small, one can easily find that the probability density of 
the reflected wavepackets with the phase-shift given by Eq.~\ref{eq-phase-shift} for large times 
will take a form 
\begin{eqnarray} 
	|\psi^{q,\sigma}_{k,\Delta k}(t)|^2 \approx \frac{1}{\pi}
		\frac{\sin^2\left( x(t)\Delta k/2 \right)}{x(t)^2 \Delta k/2}
\end{eqnarray} 
where $x(t)=x+d\theta^{q,\sigma}/dk - kt$, i.e. the maximum of the probability is at 
$x=kt - d\theta^{q,\sigma}/dk$. From this it is evident that the electron with spin down will be 
lagging behind that with spin up by a distance $2 l^a=d\theta^{q,\uparrow}/dk - d\theta^{q,\downarrow}/dk$
(indicated for the reflected right-going wavepackets in Fig.~\ref{fig-1}).
This behavior of the scattering of single electron in the wavepacket can be directly extended 
to many-electrons in view of the stroboscopic wavepacket basis\cite{Bokes08}. Namely, the 
above wavepackets for times $t_m= t + 2\pi m/(k\Delta k)$, $m=0,\pm 1, \pm 2, ... $, where $t$ 
is arbitrary moment of physical time, form orthogonal set of states which are all occupied 
with electrons. Hence the product of the length $2l^a$ times the number of electrons per area 
must give an estimate of the spin-polarization per unit length of the edge, as given 
in our previous work. Here we will show that this physically motivated estimate is in fact 
a rigorous result that we formally prove in the following, directly using the eigenstates of 
the Hamiltonian of the studied system.

The spin polarization {\it per unit length of the edge} is given in terms of the spin-resolved density as
\begin{eqnarray}
        m_z = \int_{-\infty}^{+\infty} dx \int \frac{dq}{2\pi} \int \frac{dk}{2\pi} 
	(n^{k,q}_\uparrow(x) - n^{k,q}_\downarrow(x) ), \label{eq-mz-1}
\end{eqnarray}
where the integration in $q$ and $k$ goes over all the occupied eigenstates and $n^{k,q}_\sigma(x)$ 
is the contribution to the density at position $x,y$ from an occupied eigenstate
\begin{equation}
        n^{k,q}_\sigma(x) = \left| e^{iqy} \left( e^{-ikx} + r^{q,\sigma}(k) e^{ikx} \right) \right|^2
                        = 2 + 2 \Re \left\{ e^{i(2kx + k l^{q,\sigma})} \right\}.\label{eq-eigenstates}
\end{equation}
Since the spatial shift $l^{q,\sigma}$ is very small (given by the strength of the SO interaction), we can
approximate the expression for the spin density to the first order
\begin{eqnarray}
        n^{k,q}_\sigma(x) &=& \left. n^{k,q}_\sigma(x) \right|_{l=0}  
		+ \left. \D{~}{l} n^{k,q}_\sigma(x) \right|_{l=0} l^{q,\sigma}, \\
		&=& \left. n^{k,q}_\sigma(x) \right|_{l=0} 
		+ \left. \frac{1}{2}\D{~}{x} n^{k,q}_\sigma(x) \right|_{l=0} l^{q,\sigma} \label{eq-mz-3}
\end{eqnarray}
where we have interchanged the differentiation in view of the functional dependence
of the density, Eq.~\ref{eq-eigenstates} on $x$ and $l^{q,\sigma}$.
Hence, using Eq.~\ref{eq-mz-1} and Eq.~\ref{eq-mz-3} we obtain for the spin polarization
\begin{eqnarray}
 	 m_z &=& \int \frac{dq}{2\pi} \int \frac{dk}{2\pi}
                \left. n^{k,q}_\uparrow(x) \right|_{x=-\infty}^{x=+\infty} l^{q}_a, \label{eq-mz-4}
\end{eqnarray}
where we have introduced the antisymmetric part of the spatial shift 
\begin{eqnarray}
l^q_a = \frac{\theta^{q,\uparrow} - \theta^{q,\downarrow}}{2k}= 
		 - 2 \alpha_E q 
		- 4 \alpha_E^3 q^3 (\Delta V - \frac{2}{3} k^2).  \label{eq-l-a} 
\end{eqnarray} 
The density at $x=-\infty$ is exponentially small, 
where as the contribution in the bulk of the 2DEG is according to Eq.~\ref{eq-eigenstates}
equal to $2$. 

Including higher order terms in Eq.~\ref{eq-mz-3} would result in appearance of contributions  
with higher spatial derivatives of the density at $x=\pm \infty$. There, however, is the density 
constant (either zero for $x=-\infty$ or the homogeneous 2DEG value for $x=+\infty$) so that the 
expression for the spin polarization, Eq.~\ref{eq-mz-4} is correct also for large values of $l^{q,\sigma}$
in principle.

To complete the derivation we need to integrate over the occupied states. The simplest way is to 
assume that the non-equilibrium distribution is well described by a Fermi sphere with the radius (Fermi momentum) $k_F$ and the corresponding Fermi energy $E_F=k_F^2/2$ in 2D, shifted by a drift momentum $q_d$ 
in the $+y$ direction. Such a model corresponds to the distribution 
function maximizing the information entropy for a fixed average energy, density 
and current~\cite{Bokes03,Bokes05}, and is also supported by Boltzmann-equation based analysis~\cite{Rech09}.
In the Appendix B we discuss the results obtained using the non-equilibrium model with two 
Fermi energies, conventional within the meso- and nano-scopic quantum transport. It gives identical 
result for the first order expansion in terms of $\alpha_E$; higher order terms in $\alpha_E$ differ 
by a numerical prefactor but are negligibly small for any realistic situations.

Using the expression for the asymmetric spatial shift, Eq.~\ref{eq-l-a}, we find for 
the induced spin polarization linear in the drift momentum (and hence the current density)
\begin{eqnarray}
         m_z &=& \int_{-k_F}^{k_F} \frac{dq}{2\pi} \int_{0}^{\sqrt{2E_F-q^2}} \frac{dk}{\pi} \left[ 
		l^{q+q_d}_a \right] \nonumber \\
	     &=& \left(  - \alpha_E  - 3 \alpha_E^3 \Delta V E_F  + \frac{2}{3} \alpha_E^3 E_F^2 
		\right) q_d n_{2D}
\end{eqnarray}
where $k_F=\sqrt{2E_F}$ is the Fermi momentum, $n_{2D}=E_F/\pi$ is the density of the 2DEG.
(See Fig.~\ref{fig-app-1} for the explanation of the integration domain for $k, q$.)
On the other hand, the current density is $j=q_d n_{2D}$ so that we find the final result
\begin{equation}
	m_z = - \alpha_E j - 3 \alpha_E^3 \Delta V E_F j + \frac{2}{3} \alpha_E^3 E_F^2 j
                + \mathcal{O}(\alpha_E^5,j^3) \label{eq-m_z-6}
\end{equation}

In a typical situation, all except the first term of this expansion are negligibly small so that 
the resulting spin polarization is independent of the height of the confining potential 
$\Delta V$ and, according to the numerical results presented in the following section, 
practically independent of other gentle modifications of the model so that the first term 
should be useful as a general estimate of the magnitude of the current induced spin polarization
in heterostructures based on 2DEG. Expressing the first order result in the S.I. units we 
obtain the central result of our work, stated in the introduction in Eq.~\ref{eq-m_z-SI}.

\section{Effects of the finite slope of the edge and the partial self-consistency}
\label{sec-4} 

The aim of this section is to explore the rigidity of the result in Eq.~\ref{eq-m_z-SI} with respect 
to changes in the model of the confining potential at the edge. We will first address the dependence 
of our result on finite slope of the edge potential on the distance $d$ using the potential energy
\begin{equation} 
	V_d(x) = \left\{ \begin{array}{cc} 
			\Delta V & x < 0 \\
			\Delta V - (\Delta V/d) x & 0 < x < d \\
			0   & x > d 
			\end{array} \right. \label{eq-Vd-1}
\end{equation}
Qualitatively we can expect this dependence to be similarly weak as the dependence 
on $\Delta V$ obtained in the previous section (Eq.~\ref{eq-m_z-6}). The reason behind this 
is a mutual cancellation of two effects: (1) increasing $\Delta V$ increases the strength of the SO 
coupling in the Hamiltonian and hence the spin polarization, (2) increasing $\Delta V$ decreases 
the amplitude of the wavefunctions in the region of the nonzero gradient of the confining 
edge potential and hence its sensitivity to the SO coupling. 

To characterize this dependence quantitatively we have evaluated the expression 
Eq.~\ref{eq-mz-1} numerically, where the spin-dependent contribution to the density, $n^{k,q}_{\sigma}(x)$,
was obtained from the eigenfunctions of the effective 1D Hamiltonian containing the potential energy 
Eq.~\ref{eq-Vd-1} and, according to Eq.~\ref{eq-V-SO-2D}, from it derived spin-orbit potential energy.
The eigenfunctions were obtained by matching the plane-waves in regions $x<0$ and $x>d$ with 
the Airy functions in the region of the potential energy with linear slope, $0<x<d$. 
The integration over $k$ in Eq.~\ref{eq-mz-1} could be done analytically, whereas the integral over $q$ was 
done numerically on the regular mesh with $N_k$ points.
We used model parameters that are typical for 2DEG created within GaAs quantum wells~\cite{Sih05},
electronic density $n=0.985$ corresponding to the Fermi energy $E_F=3.01$, the current density 
$j=0.1$ and the value of the spin-orbit parameter $\alpha_E=5.53 \times 10^{-4}$.

\begin{figure}
\begin{center}
      \includegraphics[width=8.5cm]{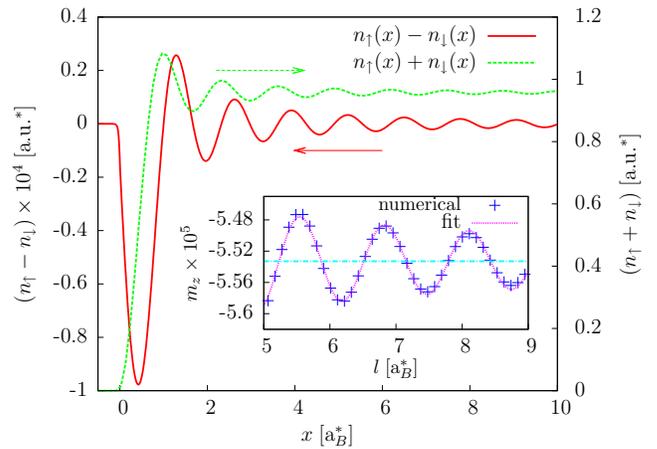}
\caption{(Color online) The difference in spin densities (red-full) and the density (green-dashed) 
	of electrons close to the edge of the sample (located at $x=0$). The former, similarly 
	to the density, exhibits Friedel-like oscillations. Its first dominant local minimum 
	(at $x \approx 0.2$) dominates the contribution to the nonzero spin-polarization obtained 
	by integrating the difference in spin densities over whole $x$ axis. The inset demonstrates 
	the extrapolation of this integration with respect to its upper limit, $l$, using a fit 
	to a functional form $\sim A\sin(2k_F l+ \phi)/l$.} \label{fig-2}
\end{center}
\end{figure}

For $d=0$ we confirm our analytical results for abrupt step potential, Eq.~\ref{eq-m_z-6}. The 
difference between the spin-up and spin-down 
densities, shown in Fig.~\ref{fig-2}, exhibits Friedel-like oscillations with the first 
period having a pronounced negative amplitude giving the major contribution to the non-zero 
spin polarization. The proportion of this amplitude to the electron is $\sim 10^{-4}$ for the current 
density $j=0.1$. Since the current density in the experimental situation is typically $j=0.1-0.01$
(the amplitude scales linearly with the current density for these values of $j$), this value 
is comparable to the amplitude of the spin polarization in the extrinsic spin Hall effect\cite{Sih05b}. 
Unfortunately, in contract to the extrinsic spin Hall effect, the nonzero spin-density 
is located only within few Fermi wavelengths away from the edge. 

Integrating the difference in the spin densities along the whole $x$ axis we obtain 
the spin-polarization per unit length of the sample. The numerical result of this integration 
naturally depends on the upper limit of the integration; a reliable result, is easily obtained 
extrapolating the upper limit to infinity (inset of Fig.~\ref{fig-2}).
Using this methodology we find the numerical result for the spin polarization 
$m_z=(5.53 \pm 0.003 ) \times 10^{-5}$, which is in perfect agreement with our analytical 
expression, Eq.~\ref{eq-m_z-6}. 

Using this extrapolation scheme we can address the changes in the spin-polarization with parameters 
of the model discussed in the following. First, we confirm the observation that the dependence 
on the magnitude of the confining edge potential comes in the 3rd order of $\alpha_E$: the value 
of the slope of the polarization vs. the height of the confinement edge potential, $\Delta V$ is 
$\partial m_z/\partial (\Delta V) \approx 1.6 \times 10^{-10}$ which is close to the analytical value $1.5 \times 10^{-10}$.
The difference comes primarily from the extrapolation of the value of $m_z$ with respect 
to the upper limit of integration in $x$. 

Considering nonzero values of the smearing length of the confining potential, $d \in (0,\lambda_F/2)$, 
we find that the resulting spin polarization is changing only very little ($\partial m_z/\partial d \sim 
3.3 \times 10^{-8}$). The reason for this negligible dependence is, similarly to the dependence 
on $\Delta V$,  the mutual cancellation of the two effects discussed earlier.

\begin{figure}
\begin{center}
      \includegraphics[width=8.0cm]{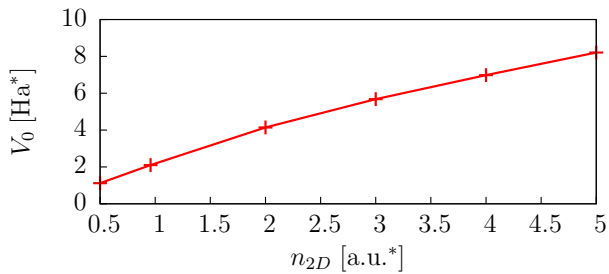}
\caption{(Color online) The dependence of the self-consistent value of the potential dip $V_0$ on 
	the electronic density shows monotonic behavior. While the value of the dip is 
	comparable to the Fermi energy ($E_F=3.01$), the considered values do not lead to appearance 
	of the bound state at the edge.} \label{fig-V0-n2DEG}
\end{center}
\end{figure}

The second modification to the confining potential that we consider is partial self-consistency
of the edge potential that guarantees charge neutrality of the edge of the sample. 
Presence of the model confining edge (Eq.~\ref{eq-Vd-1}) leads to redistribution 
of the charge density characterized with excess positive charge (per unit length of the sample)
\begin{equation} 
	\Delta N = \int_{-\infty}^{+\infty} dx \Delta n(x) = \int_{-\infty}^{+\infty} dx (E_F/\pi - n(x)),
\end{equation} 
where $n(x)$ is the electronic density and $E_F/\pi$ is the density of the positive background charge. 
The Fermi energy of the electrons guarantees that for large $x$ we have
$n(x)=E_F/\pi$ so that the system is charge neutral in the bulk even through the total charge 
at the edge is not necessarily zero. The behavior of the density at the edge of a 2D sample is in contrast 
with the typical situation for the edge of 3D metals where the work function and the Fermi energy, 
mutually comparable in value, lead to leakage of the electronic density into vacuum and thereby to 
an access of negative charge~\cite{LiebschBook}. However at the edge of a 2DEG, the Fermi energy, 
typically few meV is much smaller than the work function which is of order of several eV 
and the situation is reversed.

The overall charge of the edge can be neutralized by a simple model potential
\begin{equation} 
	V_n(x) = \left\{ \begin{array}{cc} 
			\Delta V & x < 0 \\
			-V_0 + (2 V_0/\lambda_F) x & 0 < x < \lambda_F/2 \\
			0   & x > \lambda_F/2 
			\end{array} \right. \label{eq-Vn-1}, 
\end{equation}
which on a distance $\lambda_F/2$ exhibits small potential dip to attract more electrons. 
The magnitude of the dip is obtained self-consistently to achieve zero total charge at the edge, 
i.e. the condition $\Delta N = 0$. While in principle this form of the potential may support new 
bound states (edge states) for sufficiently large $V_0$, we have checked that for the here, 
self-consistently found values of $V_0$ no such states exist. The dependence of the self-consistent 
value of $V_0$ on the typical values of the density of the 2DEG is shown in Fig.~\ref{fig-V0-n2DEG}.

\begin{figure}
\begin{center}
      \includegraphics[width=8.5cm]{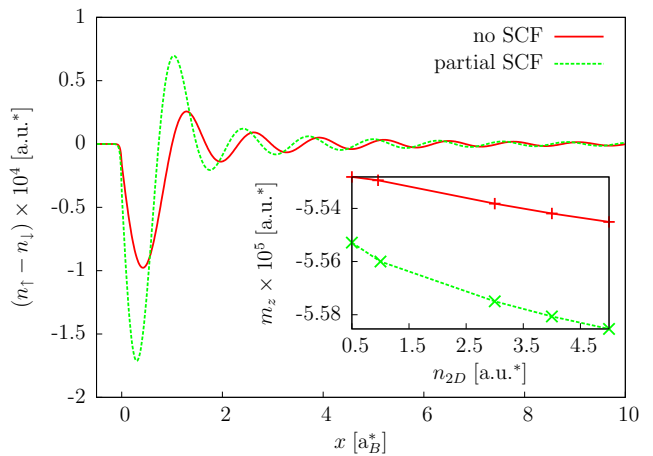}
\caption{(Color online) The difference in the spin densities for non-interacting (red-full) 
	and the partially self-consistent (green-dashed) electrons close to the edge of the sample 
	(located at $x=0$). The spatially resolved spin density difference exhibits pronounced differences 
	due to the selfconsistency, the most important effect being the shift of the curve 
	closer to the edge. 
	However, this results only in weak enhancement of the total spin polarization, $m_z$, 
	shown in the inset for an interval of electronic densities, due to counter-acting spin-orbit 
	coupling induced by the self-consistency correction in the region $x\in(0,\lambda_F/2)$.
	} \label{fig-4}
\end{center}
\end{figure}

Including the SO interaction, Eq.~\ref{eq-V-SO-2D}, on top of the potential energy $V_n(x)$,
we obtain aditional repulsive SO-induced contribution for the spin-down electrons due to the linearly 
rising potential of the dip for $0<x<\lambda_F/2$. This is in competition with the character 
of the edge-confinement potential preferring the spin-down electrons. However, the dip in 
the potential energy also tends to move the electrons closer to the confining edge 
and, by increasing the density in this region, effectively enhances the effect of the SO coupling 
there. While the form of the spin-density does change significantly due to all these mechanisms 
(see Fig.~\ref{fig-4}), the overall spin polarization per unit length remains essentially unaltered 
over a wide range of electronic densities, shown in the inset of Fig.~\ref{fig-4}. This result once again 
confirms the  rigidity of the result given by Eq.~\ref{eq-m_z-SI}.

\section{Conclusions}
\label{sec-5}

The confining potential close to the edge of a 2D electron gas together with a nonzero 
current along this edge induces a nonzero spin polarization that is localized within 
a few Fermi wavelengths from the edge. To characterize this effect quantitatively we have derived 
a simple analytical formula for the spin polarization per unit length of the edge. Interestingly, 
the spin polarization is independent of the height of the confining potential as well as the 
electronic density of the 2DEG. Furthermore, using numerical calculations we have also showed 
that this result is independent on other possible modifications of the shape of the confining potential: 
the spatial extent of the confining potential and the partial selfconsistency of the confining potential 
with respect to charge neutrality of the edge. 

\acknowledgements
The authors wish to acknowledge fruitful discussions with J. T\'{o}bik and
L. Ki\v{c}\'{i}nov\'{a}. P.B. would like to thank Rex Godby for many stimulating discussions.
This research has been supported by the Slovak grant agency VEGA (project No. 1/0452/09) and 
the NANOQUANTA EU Network of Excellence (NMP4-CT-2004-500198).

\section{Appendix A}
\label{sec-app-A}

The calculation of the spin-dependent phase-shift follows the usual textbook treatment of the wave-function
matching method in 1D problems. For the separated $x-$ dependent factor of the eigenstate we need to solve 
the 1D Schr\"{o}dinger equation
\begin{equation}
	\left( -\frac{1}{2} \DD{}{x} + \Delta V \theta(-x) 
	\pm \alpha_E q \Delta V \delta(x) \right) \phi^{q,\sigma}_e(x) = e \phi^{q,\sigma}_e(x), \label{eq-eff-H}
\end{equation}
where the `$+$' sign is for $\sigma=\uparrow$ spin and the '$-$' sign for $\sigma=\downarrow$ spin.
We are only interested in the solution well below the vacuum, i.e. $e< \Delta V$ for which we demand 
asymptotic forms
\begin{eqnarray}
        \phi^{q,\sigma}_e(x) = t^{q,\sigma}(\kappa) e^{\kappa x}, \quad x<0, \kappa=\sqrt{2(\Delta V - e)}  \\
        \phi^{q,\sigma}_e(x) = e^{-ikx} + r^{q,\sigma}(k) e^{ikx}, \quad x>0, k= \sqrt{2e}
\end{eqnarray}
The reflection amplitude ($r^{q,\sigma}(k)$) and the coefficient $t^{q,\sigma}(\kappa)$ are then found from the
continuity of $\phi^{q,\sigma}_e(x)$ and its derivative at $x=0$ with the result
\begin{eqnarray}
        r^{q,\sigma}(k) &=& - \frac{\pm 2\alpha_E q \Delta V +  \kappa + ik}{\pm 2 \alpha_E q \Delta V 
		+ \kappa - ik} \\
        t^{q,\sigma}(\kappa) &=&  - \frac{2ik}{\pm 2 \alpha_E q \Delta V + \kappa + ik}.
\end{eqnarray}
The sought phase shift is obtained from the reflection amplitude 
$r^{q,\sigma}_e = |r^{q,\sigma}(k)|e^{i\theta^{q,\sigma}_k}$ ,
\begin{eqnarray} 
	\theta^{q,\sigma}_k = \pi + 2\textrm{atan}\frac{k}{\pm 2 \alpha_E q \Delta V + \kappa}.
\end{eqnarray}
Expanding the last result in powers of $\alpha_E$ to $3^{rd}$ oder we obtain the expression 
Eq.~\ref{eq-phase-shift}.

\section{Appendix B}
\label{sec-app-B}

\begin{figure}
\begin{center}
      \includegraphics[width=7cm]{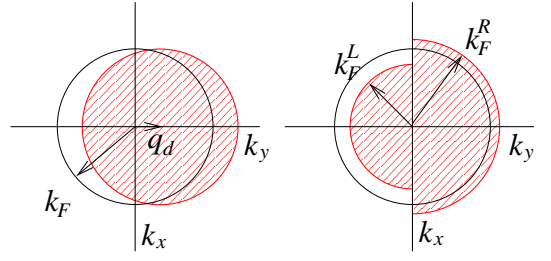}
\caption{The shifted Fermi sphere (left) and the two-Fermi radii (right) occupations in 2DEG.
	The occupied states are shown as filled areas, have both identical current density and electronic 
	density but differ in higher moments of $k$.} \label{fig-app-1}
\end{center}
\end{figure}

We have stated that the result for the spin-polarization, Eq.~\ref{eq-m_z-6}, 
is independent of the considered non-equilibrium occupations of electronic states to the first order in 
the the spin-orbit coupling $\alpha_E$. This is true as long as the current density in the system is the same 
for all of these considered occupations. Let us consider two simple models of occupations: (1) 
the ``two-Fermi radii model'' (2FR), frequently used within the coherent transport with the occupancies 
dictated by the left and right macroscopic electrodes with Fermi momenta $k_F^R$ and $k_F^L$ 
respectively~\cite{Mera05} attached to the sample at its ends (Fig.~\ref{fig-app-1}, right), 
and (2) the shifted Fermi distribution function (ShF), used within the main text (Fig.~\ref{fig-app-1}, left)
From this one immediately sees that the contribution to the spin polarization that is linear in 
$\alpha_E$ in Eq.~\ref{eq-mz-4}, and therefore also linear in the momentum $k_y (=q)$, will be independent 
of the particular form of the occupations as long as the zero-th and the first moments are identical.

Apart from this, the 2FR and ShF occupations represent two different situations: the former 
is suited for very short ballistic systems and it directly facilitates interpretation of the results 
in terms of the difference in electro-chemical potentials of the electrodes,
$\Delta \mu = (1/2)\left( (k_F^L)^2 - (k_F^R)^2 \right)$. The latter is related to the current 
density inside the sample through 
\begin{eqnarray}
	j &=& \int \frac{dk_x dk_y}{4\pi^2} n(k_x,k_y) k_y 
		= \frac{1}{3\pi^2} \left( (k_F^R)^3 - (k_F^L)^3 \right)  \\
	&\approx& \frac{1}{\pi^2} \sqrt{2E_F} \Delta \mu, \label{eq-j-mu-lin}
\end{eqnarray}
where the occupation factor, $n(k_x,k_y)=2$ (spin degeneracy in the unperturbed system) 
for the occupied states shown in Fig.~\ref{fig-app-1} and $n(k_x,k_y)=0$ otherwise.  
Eq.~\ref{eq-j-mu-lin} gives the linear expansion in the applied bias $\Delta \mu$.
No such simple connection to the applied bias voltage can be made for the ShF occupation.
On the other hand, ShF is suitable for longer samples, where electrons' scattering leads 
to partial equilibration of the distribution function~\cite{Rech09}. This is then characterized 
with the drift momentum, $q_d$, which gives the current density 
\begin{equation} 
	j = n_{2D} q_d = \frac{E_F}{\pi} q_d.
\end{equation}
Since the samples we consider are typically longer than the electron's coherence length, we have used 
the ShF occupations within the main text. For completeness we give also the results for the 2FR case.
The spin polarization per unit length of the sample to third order in $\alpha_E$ is
\begin{equation}
	m_z = - \frac{\alpha_E }{\pi^2} \sqrt{2E_F} \Delta \mu 
		- \frac{2\alpha_E^3 \Delta V}{3 \pi^2} E_F^{3/2}
		\Delta \mu - \frac{4\alpha_E^3}{45 \pi^2} E_F^{5/2} \Delta \mu
\end{equation}
which after expressing the applied bias in terms of the current density results in
\begin{equation}
	m_z =  - \alpha_E j - \frac{2\alpha_E^3 \Delta V}{3} E_F j
                + \frac{4\alpha_E^3}{45} E_F^{2} j.
\end{equation}
Comparing the last equation with Eq.~\ref{eq-m_z-6} we see that while the first order  term
is identical for both occupations, the higher orders differ by a numerical prefactor.
Both of these are larger in the case of partially equilibrated shifted Fermi-like occupations.

\bibliographystyle{prsty}

\end{document}